\documentclass{aa}
\usepackage{graphicx}
\usepackage{txfonts}
\def\nab{\mbox{\boldmath $\nabla$}}

\begin{document}


\title{On the role of meridional flows in flux transport dynamo models}

\author{Jouve Laur\`ene \& Brun Allan Sacha}

\offprints{Jouve Laur\`ene}

\institute{Laboratoire AIM, CEA/DSM-CNRS-Universit\'e Paris Diderot, DAPNIA/SAp, 91191 Gif sur Yvette, France\\
              \email{ljouve@cea.fr}}

\date{Received 9 January 2007 / Accepted 5 June 2007}

\abstract
{The Sun is a magnetic star whose magnetism and cyclic activity is linked to the existence of an internal dynamo.}
{We aim to understand the establishment of the solar magnetic 22-yr cycle, its associated butterfly diagram and field parity selection through numerical simulations of the solar global dynamo. Inspired by recent observations and 3D simulations that both exhibit multicellular flows in the solar convection zone, we seek to characterise the influence of various profiles of circulation on the behaviour of solar mean-field dynamo models. We focus our study on a number of specific points: the role played by these flows in setting the cycle period and the shape of the butterfly diagram and their influence on the magnetic field parity selection, namely the field parity switching from an antisymmetric, dipolar field configuration to a symmetric, mostly quadrupolar one, that has been discussed by several authors in the recent literature.} {We are using 2-D mean field flux transport Babcock-Leighton numerical models in which we test several types of meridional flows: 1 large single cell, 2 cells in radius and 4 cells per hemisphere.} { We confirm that adding cells in latitude tends to speed up the dynamo cycle whereas adding cells in radius more than triples the period. We find that the cycle period in the four cells model is less sensitive to the flow speed than in the other simpler meridional circulation profiles studied. Moreover, our studies show that adding cells in radius or in latitude seems to favour the parity switching to a quadrupolar solution.}{According to our numerical models, the observed 22-yr cycle and dipolar parity is easily reproduced by models including multicellular meridional flows. On the contrary, the resulting butterfly diagram and phase relationship between the toroidal and poloidal fields are affected to a point where it is unlikely that such multicellular meridional flows persist for a long period of time inside the Sun, without having to reconsider the model itself.} 

 
\maketitle

%
\section{Introduction}

The Sun possesses striking magnetic and dynamical properties, such as its turbulent convective envelope, large-scale surface differential rotation, 22-yr cycle of magnetic activity, butterfly diagram of sunspot emergence, hot corona, etc. (Stix \cite{Stix1}). 
Understanding how the physical processes operating in the solar turbulent plasma nonlinearly interact to yield this wide range of 
dynamical phenomena is very challenging. 
One successful and powerful approach is to rely on multi-dimensional magnetohydrodynamics (MHD) numerical simulations.
Today, despite tremendous advances in building powerful supercomputers, it is still not possible to
compute a fully integrated 3-D MHD model of the Sun starting from its core up to its corona. One is thus forced to study individually complementary pieces of the full solar MHD puzzle and to progressively incorporate them in a more nonlinearly coupled model.  One important characteristic of the Sun that needs to be understood is the origin of its magnetic activity because it has direct societal impact by impairing satellites, damaging electric power grids, interfering with high frequency radio communications and radars.  
It is currently believed that the solar magnetism is linked to an internal dynamo (Parker \cite{Parker1}). 
More precisely the Sun is the seat of both a small scale and irregular dynamo and a large scale and cyclic dynamo that generate 
and maintain its magnetic field and lead to the various magnetic phenomena observed at its surface (Parker \cite{Parker2}; Cattaneo \& Hughes \cite{Cattaneo1};
Ossendrijver \cite{Ossendrijver}). Developing numerical models of the solar dynamo has thus been a very active field of research. This has mainly involved
two types of numerical experiments:
\begin{itemize}
\item kinematic solar dynamo models that solve only the induction equation in its mean field approximation and assume the velocity field as given (Steenbeck \& Krause \cite{Steenbeck}; Roberts \cite{Roberts}; Stix \cite{Stix2}; Moffat \cite{Moffat}; Krause \& Radler \cite{Krause}; see Charbonneau \cite{Charbonneau1} and Solanki et al. \cite{Solanki} for recent reviews). These models rely on the parametrization of two important effects that are thought to be at the origin of the solar global dynamo, the $\alpha$ and $\Omega$ effects. They provide a useful and fast tool to model the solar 22-yr magnetic cycle and its associated butterfly diagram since no feedback from the Laplace force on the motion is accounted for.
\item or dynamical solar dynamo models 
that solve explicitly the full set of MHD equations (Gilman \cite{Gilman}; Glatzmaier \cite{Glatzmaier}; Cattaneo \cite{Cattaneo2}; Brun et al. \cite{Brun1}). These models self-consistently compute all the physical processes in
three dimensions allowing significant progress to be made on the intricate interactions operating in a turbulent magnetized plasma. The cost of 3D models and the large number of degrees of freedom needed to model the whole Sun make it difficult, as of today, to provide quantitative predictions such as the cycle period.
\end{itemize}
 Clearly, both approaches are complementary and are needed to better understand the magnetic solar activity.
Since the original ideas of Parker regarding the operation of a hydromagnetic dynamo in the Sun, many articles have 
been written to improve our understanding of this subtle physical process. 
In the late 70's, solar dynamo models were relying on a cylindrical differential rotation profile and an $\alpha$-effect linked to
non-reflexion symmetric motions within the turbulent and rotating solar convection zone (CZ), the so called $\alpha-\Omega$ dynamo.
In such dynamo models, the product of $\alpha$ and
$\partial \Omega/\partial r $ must be negative in the northern hemisphere in order to obtain an equatorward butterfly diagram (Yoshimura \cite{Yoshimura}).
However, these distributed $\alpha-\Omega$ dynamo models have since been discarded for two main reasons: first
the inversion in the mid 80's of the internal solar rotation profile  (Brown et al. \cite{Brown}; Thompson et al. \cite{Thompson}) showed a conical differential rotation profile in the convection zone  ($\partial\Omega/\partial r \simeq 0$) rather than a cylindrical profile. Secondly, it was demonstrated that strong magnetic fields could significantly reduce the efficiency of the $\alpha$-effect, in a phenomenon called $\alpha$-quenching (Ossendrijver \cite{Ossendrijver}). 
In a landmark paper, Parker (\cite{Parker2}) proposed the segregation of sites of generation of the poloidal field on the one hands with that of the toroidal field in the other hand, 
in what is now called the {\sl interface dynamo}.
He was encouraged by the latest helioseismic inversions which indicated the existence of 
a swift transition from the differential rotation of the solar convection zone to an inner solid body rotation in the radiative 
interior, i.e the tachocline (Spiegel \& Zahn \cite{Spiegel}). In the late 90's, Charbonneau \& Mc Gregor (\cite{Charbonneau2}), were the first to incorporate
all the ingredients of the modern interface dynamo: a solar-like (conical) differential rotation + a tachocline, a separate site
of generation of poloidal field (in the convection zone) vs the toroidal field (in the tachocline). They showed that with this new
solar dynamo model, the 22-yr cycle period, the butterfly diagram, the phase relationship between the poloidal and toroidal fields and the field parity can be reproduced. However, these models do not include meridional circulation (MC). To address this issue, Dikpati \& Charbonneau (\cite{Dikpati1}) computed Babcock-Leighton (BL) models (Babcock \cite{Babcock}; Leighton \cite{Leighton}; Choudhuri et al. \cite{Choudhuri}) with a solar-like $\Omega$ profile and an unicellular meridional flow. They showed that a solar dynamo model based on this so-called Babcock-Leighton flux transport
dynamo could also be successful at reproducing most of the solar global magnetic properties.
In this model, the meridional circulation transports the poloidal field from the surface, where it appears 
through the twisted nature of the solar active regions, to the bottom of the convection zone where it is transformed into a toroidal field
in the tachocline. This meridional circulation thus plays a major role in the behaviour of BL flux transport dynamo models. It is then important to understand its origin and structure in the Sun. 

An analysis of the governing equations tells us that mean meridional flows arise from a combination of buoyancy forces, Reynolds stresses, latitudinal pressure gradients and Coriolis forces acting on the mean zonal flow (differential rotation) (Miesch \cite{Miesch2}). The competition of these physical processes make it difficult to anticipate the meridional flow profile. Inside the solar envelope, this flow is much weaker than the differential rotation, making it relatively difficult to measure. Furthermore, although it can in principle be probed by global helioseismology, its effect on global acoustic waves is weak and difficult to distinguish from rotational and magnetic effects. Thus, we must currently rely on surface measurements and local helioseismology. The $15 \,{\rm m.s^{-1}}$ poleward flow observed at the surface (Hathaway \cite{Hathaway}) has been confirmed by local helioseismology with great accuracy down to $r/R_{\sun}=0.95$ (Haber et al. \cite{Haber}) and some attempts have been made to probe the MC down to $r/R_{\sun}=0.85$ (Giles et al. \cite{Giles}; Schou \& Bogart \cite{Schou}; Braun \& Fan \cite{Braun}) but the pattern and localisation of the equatorward return flow is still not well established.
Today, the favoured solar dynamo models are of flux transport type, assuming both a source of poloidal field at the surface (a BL 
source term) and at the bottom ($\alpha$-effect like) (Bonanno et al. \cite{Bonanno1}; Dikpati et al. \cite{Dikpati2}; K{\"u}ker et al. \cite{Kuker}; Chatterjee et al. \cite{Chatterjee}). In particular, recently these models have
been successful at reproducing a series of solar cycle and even for starting to predict the next/starting solar cycle (cycle 24) (Dikpati \& Gilman \cite{Dikpati3}).

In this paper we will follow the kinematic approach, by computing 2-D axisymmetric mean field solar dynamo models of the
flux transport BL type. We seek to answer the simple following 
questions:  What is the role of meridional flows in setting the solar cycle period and butterfly diagram? 
Can the presence of multicellular meridional flows lead to variations of the general properties of the solar activity?
The motivation behind these questions is that both observational evidence via local helioseismology technics (Haber et al. \cite{Haber}) 
and 3-D MHD numerical models as described above (Miesch et al. \cite{Miesch}; Brun \& Toomre \cite{Brun0}; Brun et al. \cite{Brun1}; Browning et al. \cite{Browning}) exhibit multicellular flow both in radius and 
latitude. If such permanent multicellular flow were indeed acting continuously in the Sun, it is likely that it will lead to a
different solar global dynamo model since today most models rely on a single monolithic meridional flow to transport poloidal field from the surface down to the tachocline at the base of the solar convection zone.

The paper is organized as follows: in Sect. 2, we present the mean field induction equation and the ingredients of the model, in Sect. 3, we discuss the results of our study, mainly the effect of introducing many meridional cells both in latitude and radius 
in the model. In Sect. 4 we discuss the influence of the more complex meridional flow in setting the field parity (i.e either dipolar or quadrupolar) and we conclude in Sect. 5. Finally, the numerical techniques 
used to solve the induction equation and the boundary conditions introduced to compute the temporal evolution of our solar dynamo models are presented in the appendix.

\section{Setting the solar dynamo model}

\subsection{Mean field equations}

To model the solar dynamo, we use the hydromagnetic induction equation, governing the evolution of the large scale magnetic field ${\bf B}$ in response to advection by a flow field ${\bf v}$ and resistive dissipation.

$$
\frac{\partial {\bf B}}{\partial t}=\nabla\times ({\bf v} \times{\bf B})-\nabla\times(\eta\nabla\times{\bf B}) 
$$
where $\eta$ is the effective magnetic diffusivity.

Working in spherical coordinates and under the assumption of axisymmetry, we write the total mean magnetic field {\bf B} and the velocity field {\bf v} as:
$$
{{\bf B}}(r,\theta,t)=\nab\times (A_{\phi}(r,\theta,t) \hat {\bf e}_{\phi})+B_{\phi}(r,\theta,t) \hat {\bf e}_{\phi}
$$
$$
{{\bf v}}(r,\theta)={\bf v_{p}}(r,\theta) + r\sin\theta \, \Omega(r,\theta) \hat {\bf e}_{\phi}
$$

Note that our velocity field is time-independant since we will not assume any fluctuations in time of the differential rotation $ \Omega$ or of the meridional circulation ${\bf v_{p}}$.
Reintroducing this poloidal/toroidal decomposition of the field in the mean induction equation, we get two coupled partial differential equations, one involving the poloidal potential $A_{\phi}$ and the other concerning the toroidal field $B_{\phi}$. 

\begin{equation}
\label{eqA2}
\frac{\partial {A_{\phi}}}{\partial t}=\frac{\eta}{\eta_{\rm t}} (\nabla^{2}-\frac{1}{\varpi^{2}})A_{\phi}-R_{e}\frac{\bf{v}_{p}}{\varpi}\cdot\nabla(\varpi A_{\phi})+C_{\rm s}S(r,\theta,B_{\phi})
\end{equation}

\begin{eqnarray}
\label{eqB2}
\frac{\partial {B_{\phi}}}{\partial t}&=&\frac{\eta}{\eta_{\rm t}} (\nabla^{2}-\frac{1}{\varpi^{2}})B_{\phi}+\frac{1}{\varpi}\frac{\partial(\varpi B_{\phi})}{\partial r}\frac{\partial (\eta/\eta_{\rm t})}{\partial r}-R_{e}\varpi {\bf v}_{\rm p}\cdot\nabla(\frac{B_{\phi}}{\varpi})\nonumber  \\ 
 & & \nonumber \\
&-&R_{\rm e}B_{\phi}\nabla\cdot{\bf v}_{\rm p}+C_{\Omega}\varpi(\nabla\times(A_{\phi}{\bf \hat{e}}_{\phi}))\cdot\nabla\Omega
\end{eqnarray}

where $\varpi=r\sin\theta$, $\eta_{\rm t}$ is the turbulent magnetic diffusivity (diffusivity in the convective zone), ${\bf v}_{\rm p}$ the flow in the meridional plane (i.e. the meridional circulation), $\Omega$ the differential rotation. The break of axisymmetry needed to circumvent Cowling's anti-dynamo theorem comes from the addition of a term $S(r,\theta,B_{\phi})$ in Eq.(\ref{eqA2}), representing the BL surface source term for poloidal field.
In order to write these equations in a dimensionless form, we choose as length scale the solar radius $R_{\sun}$ and as time scale the diffusion time $R_{\sun}^2/\eta_{\rm t}$ based on the envelope diffusivity $\eta_{\rm t}$.
This leads to the appearance of three control parameters $C_{\Omega}=\Omega_{0}R_{\sun}^2/\eta_{\rm t}$, $C_{\rm s}=s_{0}R_{\sun}/\eta_{\rm t}$ and $R_{\rm e}=v_{0}R_{\sun}/\eta_{\rm t}$ where $\Omega_{0}, s_{0}, v_{0}$ are respectively the amplitude of the differential rotation, of the surface source term and of the meridional flow.

Equations $\ref{eqA2}$ and $\ref{eqB2}$ are solved in an annular meridional cut with the colatitude $\theta$ $\in [0,\pi]$ and the radius $r \in [0.6,1]R_{\sun}$ i.e from slightly below the tachocline ($r=0.7R_{\sun}$) up to the solar surface, using a finite element method (STELEM code) which was validated thanks to an international dynamo benchmark (Jouve et al. \cite{Jouve}, see the appendix for more details on the numerical technique). At $\theta=0$ and $\theta=\pi$ boundaries, both $A_{\phi}$ and $B_{\phi}$ are set to 0 and at $r=0.6R_{\sun}$, both $A_{\phi}$ and $B_{\phi}$ are set to $0$ . At the upper boundary,  we smoothly match our solution to an external potential field, i.e. we have vacuum for $r \geq R_{\sun}$.
As initial conditions we are setting a confined dipolar field configuration, i.e the poloidal field is set to $\sin\theta / r^{2}$ in the convective zone and to $0$ below the tachocline whereas the toroidal field is set to $0$ everywhere.

\subsection{The physical ingredients}

The model ``ingredients'' are basically those used by Dikpati \& Charbonneau (\cite{Dikpati1}).
The rotation profile is a representation of that deduced from helioseismic inversions, assuming a solid rotation below $0.65R_{\sun}$ and a differential rotation above the interface. With this profile, the radial shear is maximal in the tachocline:

\begin{eqnarray}
{\Omega(r,\theta)}=\Omega_{\rm c}+\frac{1}{2}  \left(\Omega_{\rm Eq} +a_{2}\cos^2\theta +a_{4}\cos^4\theta-\Omega_{\rm c}\right) 
\nonumber \\ \nonumber
\times \left[1+{\rm erf}\left(2\frac{r-r_{\rm c}}{d_{1}}\right) \right] \nonumber \\ \nonumber
\end{eqnarray}

with $\Omega_{\rm Eq}=1$,  $\Omega_{\rm c}= 0.93944$,  $r_{\rm c}=0.7R_{\sun}$, $d_{1}=0.05$,

$a_{2}=-0.136076$ and $a_{4}=-0.145713$.

\bigskip

In BL flux transport dynamo models, the poloidal field owes its origin to the twist of the magnetic field emerging at the solar surface. Thus, the source has to be confined in a thin layer just below the surface and as the process is fundamentally non-local, the source term depends on the variation of $B_{\phi}$ at the base of the convection zone. Moreover, a quenching term is introduced to prevent the magnetic energy from growing exponentially:

\begin{eqnarray}
S(r,\theta,B_{\phi})&=& \frac{1}{2}\left[1+{\rm erf}(\frac{r-r_{2}}{d_{2}})\right]\left[ 1-{\rm erf}(\frac{r-R_{\sun}}{d_{2}})\right] \nonumber \\ 
&\times&\left[1+\left({\frac{B_{\phi}(r_{c},\theta,t)}{B_{0}}}\right)^{2}\right]^{-1}\cos\theta \sin\theta B_{\phi}(r_{\rm c},\theta,t) \nonumber
\end{eqnarray}

where $r_{2}=0.95$, $d_{2}=0.01$, $B_{0}=10^5$.

\bigskip

We assume that the net diffusivity in the envelope $\eta$ is dominated by its turbulent contribution whereas in the stable zone, the value of the diffusivity has to be much weaker (we have $\eta_{\rm c}<<\eta_{\rm t}$). We smoothly match the two different constant values thanks to an error function which enables us to quickly and continuously transit from $\eta_{\rm c}=10^9 \, \rm cm^2.s^{-1}$ to $\eta_{\rm t}$ which is a variable parameter in our computations. It gives us the diffusivity function below:

$$
\frac{\eta}{\eta_{\rm t}}=\frac{\eta_{\rm c}}{\eta_{\rm t}}+\frac{1}{2}\left[1+{\rm erf}\left(2\frac{r-r_{\rm c}}{d_{1}}\right)\right]
$$

We have now set up a detailed model of the global solar dynamo, using the framework of mean field theory. One of the key ingredients of this kind of models is the meridional circulation. We now investigate the influence of complex flows on the solar dynamo and its global properties.

\section{Influence of meridional circulation on magnetic cycles}
\label{sec}

\subsection{Our reference unicellular model}

\begin{figure}[h]
   \centering
\includegraphics[width=9cm]{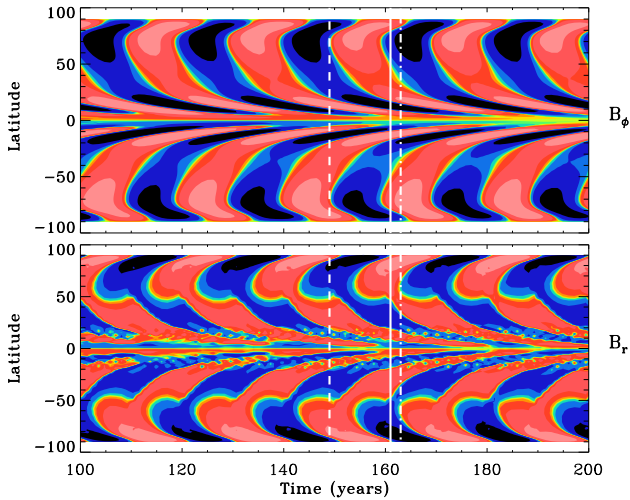}
      \caption{Reference case: butterfly diagram (time-latitude cut at $r=\rm cst$) of the unicellular model with $v_{0}=643 \, \rm cm.s^{-1}$,  $s_{0}=20 \, \rm cm.s^{-1}$ and $\eta_{\rm t}=5.10^{10} \, \rm cm^2.s^{-1}$. The contours of $B_{\phi}$ (upper panel) are plotted at the base of the convection zone and $B_{r}$ (lower panel) is taken at the surface. Contours are logarithmically spaced with 2 contours covering a decade in field strength and red colours represent positive values of the field. The vertical dashed line corresponds to the epoch of reversal of toroidal field, the plain line correspond to the epoch of reversal of poloidal field at the poles from negative to positive polarity and the dash-dotted line corresponds to the positive maximum of toroidal field near the equator.}
       \label{figure_butterflyA}
   \end{figure}

We first compute a model where we assume one large single meridional cell per hemisphere which we will consider as the reference model. The components of the meridional circulation are those used in Van Ballegooijen \& Choudhuri (\cite{VanBallegooijen}) which defines a steady circulation pattern, symmetric with respect to the equator, with a single flow cell per hemisphere directed poleward at the surface and allowed to penetrate a little below the base of the CZ, where it is equatorward. 
With this typical model, we are able to reproduce several aspects of the solar cycle, notably its period of approximately 20 years, a strong equatorward branch for toroidal field restricted to low latitudes, a phase shift of $\pi/2$ between the surface polar field and the deep toroidal field, so that the polar field changes its polarity from negative to positive when the toroidal field is positive and maximal in intensity near the equator (Fig. $\ref{figure_butterflyA}$).
Moreover, the strong equatorward branch for the toroidal field is the signature of the drag of the toroidal field by equatorward MC at the base of the convection zone and thus clearly shows the dominating effect of field advection over diffusion. Indeed, a least square fit indicates that the cycle period $T$ strongly depends on the meridional flow amplitude, i.e. ${ T\propto v_{0}^{-0.63}}$.

One could wonder why we are seeking to improve and modify the reference model given its relatively good agreements with observations. There are in fact several reasons.

First, a 26-year interval studied by Snodgrass and Dailey (\cite{Snodgrass}) exhibited large temporal variations in the meridional flow amplitude. Indeed they found that even if the latitudinal flow peaked at about $15 \, \rm m.s^{-1}$ in average, MC could achieve amplitudes as large as $50 \,\rm  m.s^{-1}$. Given the strong dependance of the cycle period on the amplitude of the flow (if we triple the velocity amplitude we reduce the period by about one-half), we can thus wonder if a systematic period of about 22 years can be conserved in this context of temporally varying flows. A study of the impact of stochasticity in such BL models (Charbonneau \& Dikpati 2000) yet showed that these solar cycle models were quite robust to stochastic variations of the meridional circulation.

Another argument against this type of BL models comes from Dikpati $\&$ Charbonneau (\cite{Dikpati1}) who showed that even if the configuration of the toroidal field seems to fit the observations quite well particularly concerning the strong equatorward branch, a relatively strong toroidal field ($10^3$ G) is also present at all latitudes. This strong field existing at all latitudes could be significantly decreased by imposing a lower threshold for quenching in the surface source term that would prevent toroidal flux tubes that are too weak in intensity to rise through the CZ and thus participate in the regeneration of the poloidal field (Charbonneau et al. \cite{Charbonneau3}). 

 Finally, Dikpati $\&$ Gilman raised in 2001 a major concern about the BL flux transport model, concerning the symmetry of the magnetic field with respect to the equator. They claim that, in the particular range of parameters they are using to get a solar-like period, the pure BL flux transport model fails to reproduce the persistent antisymmetry of the toroidal field and that whatever the magnetic initial conditions imposed, this model would always end up giving a quadrupolar configuration, which we do not currently observe in the Sun (see Sect. \ref{parity}).

Thus the single cell pure BL model does not seem fully satisfactory and needs to be improved. Moreover, both observations by Haber et al. (\cite{Haber}) and 3D simulations by Brun et al. (\cite{Brun1}) show multiple cells circulation in the CZ and modulation of the MC with magnetic fields. Dikpati et al. (\cite{Dikpati2}) and Bonanno et al. (\cite{Bonanno2}) computed dynamo models including 2 cells in latitude per hemisphere. Dikpati, with a BL source term, found that these additional cells tended to decrease the cycle period and Bonanno, with a distibuted $\alpha$-effect model, found that the global pattern of the meridional circulation could strongly influence the location of the dynamo action in the advection-dominated regime. Consequently, we would like to verify the influence of even more complex multicellular (both in radius and in latitude) meridional flow on the cycle, on the butterfly diagram, on the phase relationship between the poloidal and the toroidal parts of the magnetic field.

\begin{table*}[htbp]
\begin{center}
\caption{ Summary of the 5 different cases and associated parameters. The last column indicates the period in years for each case.}
\begin{tabular}{ccccccc} \hline 
 & {\bf  Resolution} & \bf  Time step & ${\bf v_{0}}$ & ${\bf \eta_{\rm t}}$ & ${\bf s_{0}}$ & \bf  Cycle Period \\ 
 & $n_x \times n_z$ &   & ($\rm cm.s^{-1}$) & ($\rm cm^2.s^{-1}$) & ($\rm cm.s^{-1}$) & \bf (Yrs) \\ \hline
{\bf  Reference case} & $128^2$ & $7.2 \, 10^{-7}$ & $643$ & $5.10^{10}$ & $20$ & $\bf 21.8$ \\ \hline
{\bf  Case 1a} & $256 \times 128$ & $4.53 \, 10^{-8}$ & $643$ & $5.10^{10}$ & $20$ & $\bf 84.6$ \\ \hline
{\bf Case 1b} & $256 \times 128$ & $4.53 \, 10^{-8}$ & $1916$ & $1.4910^{11}$ & $20$ & $\bf 22.4$\\ \hline
{\bf  Case 2a} & $256 \times 128$ &  $4.53 \, 10^{-8}$ & $643$ & $5.10^{10}$ & $20$ & $\bf 44.7$  \\ \hline
{\bf Case 2b} & $128^2$ & $7.2 \, 10^{-7}$ & $1071$ & $1.5 \, 10^{11}$ & $20$ & $\bf 22.4$ \\ \hline
\end{tabular}
\label{table_resume}
\end{center}
\end{table*}

\subsection{Solar dynamo models with additional cells in the meridional circulation}

We focus here on two cases, the case with 2 cells in radius and the case with 2 cells in radius and 2 in latitude. We will not deal with the 2 latitudinal cells model, as this was already treated by Dikpati et al. (\cite{Dikpati2}) and Bonanno et al. (\cite{Bonanno2}). The main characteristics of the different cases studied are summarized in Table \ref{table_resume}.

To get a multicellular flow, we write the stream function $\psi$ as a product of Chebyshev polynomials in radius and of Legendre polynomials in latitude. Through the following equations: 
 $\rho v_{r}=\frac{1}{z^2}\frac{\partial \psi}{\partial x}$ and $\rho v_{\theta}=-\frac{1}{z\sqrt{1-x^2}}\frac{\partial \psi}{\partial z}$, with $z=r$, $x=-\cos\theta$ and $\rho=1/z$, which ensure that $\nabla . (\rho {\bf v})=0$, we easily deduce the shape of the meridional flow components from the polynomial stream function. 

\subsubsection{Case 1: 2 cells in radius per hemisphere}

For this case, all the ingredients are kept identical to the reference model but the meridional flow is modified. We set the stream function $\psi$ to:

$$
\psi(x,z)=K_{1}z(z-0.6)(500(z-0.8)^3-20(z-0.8))(x^3-x)
$$

\begin{figure}[h]
\begin{center}
\includegraphics[width=8cm]{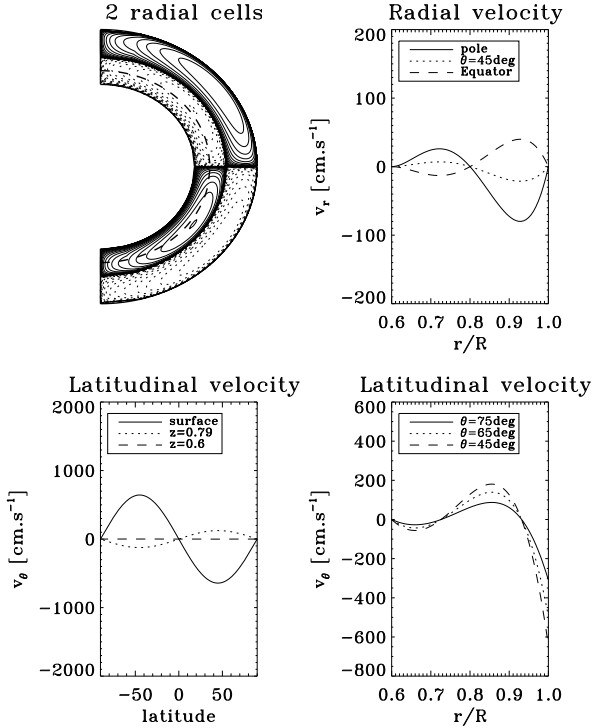}
\end{center}
\caption{Stream function and components of the meridional flow multiplied by $v_{0}=643 \, \rm cm.s^{-1}$ for the 2 radial cells model.}
\label{fig:courant2ray}
\end{figure}

$K_{1}$ is a normalization factor, i.e. it is chosen so that $v_{\theta}/v_{0}=1$ at the solar surface and at a latitude of $45 \degr$ (see Fig. $\ref{fig:courant2ray}$).

\begin{figure}[h]
   \centering
\includegraphics[width=9.5cm]{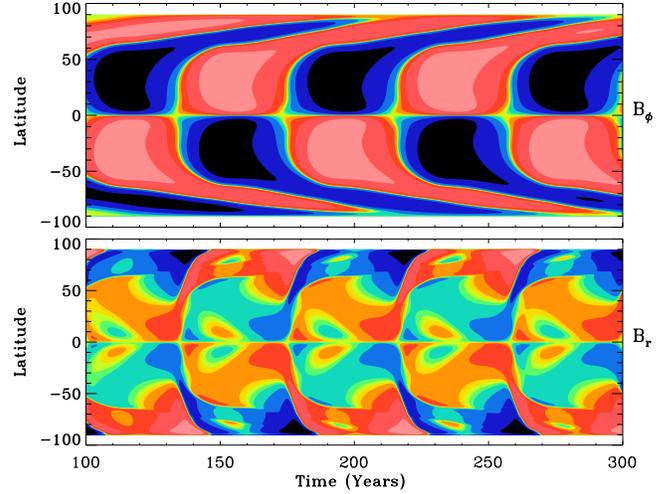}
      \caption{ Case 1a: butterfly diagram (time-latitude cut at $r=\rm cst$) of case 1 with $v_{0}=643 \, \rm cm.s^{-1}$. The format is the same as Fig. $\ref{figure_butterflyA}$. }
       \label{figure_butterfly2ray}
   \end{figure}

In Fig. $\ref{figure_butterfly2ray}$, we represent the butterfly diagram of case 1 with the parameters used in the reference unicellular model. For this model the cycle lasts 84.6 years, more than 3 times longer. The increase of the period comes from the fact that the magnetic flux is not transported from the surface to the interface as fast as it was in the unicellular model. This is a direct consequence of the presence of a return flow at mid depth. The two source regions (the surface for the poloidal field and the tachocline for the toroidal field) are thus not linked as directly as they were in the reference model. This leads to a slower regeneration of the toroidal field from poloidal field and vice versa.

\begin{figure}[h]
   \centering
\includegraphics[width=9.5cm]{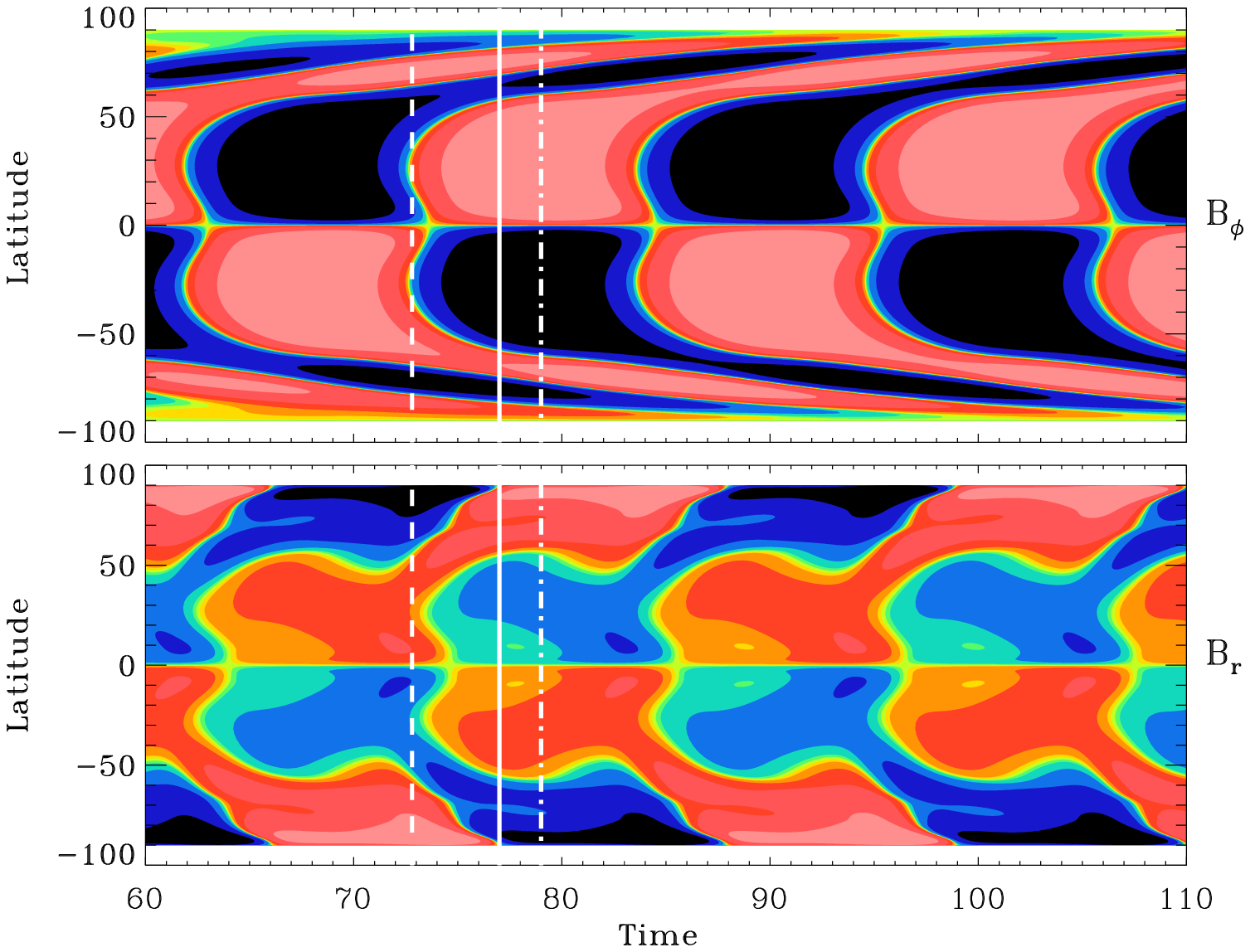}
      \caption{ Case 1b: butterfly diagram (time-latitude cut at $r=\rm cst$) of case 1b with $v_{0}=1916 \, \rm cm.s^{-1}$ and $\eta_{\rm t}=1.49\times10^{11} \, \rm cm^2.s^{-1}$. The format is the same as Fig. $\ref{figure_butterflyA}$. }
       \label{figure_dephas2ray}
   \end{figure}

We also see that since the meridional flow is directed poleward at the base of the convective zone, we get a very strong poleward branch for the toroidal field. Moreover, the last panel of Fig. $\ref{fig:courant2ray}$ shows that the latitudinal velocity intensity is 3 times higher at mid-depth (where it is equatorward) than at the base of the CZ (where it is poleward). As a consequence, the field is advected 3 times faster to the equator at mid-depth than to the poles at the base of the convection zone and this explains the strong domination over time of the poleward against the equatorward branch seen on the upper panel of  Fig. $\ref{figure_butterfly2ray}$. This figure also shows the existence of an equatorward branch between $0$ and $30 \degr$ as requested by observations. A small amount of poloidal field is thus advected towards the equator even though the flow is poleward in this region. This is a direct consequence of the non-locality of our surface source-term. Indeed, at these latitudes and between $0.73R_{\sun}$ and $0.94R_{\sun}$, the toroidal field is advected toward the equator by the MC flow. Both through advection and diffusion, this equatorward-migrating toroidal structure is transported inward near $0.7R_{\sun}$ and as the poloidal source term is linked to the toroidal field at this radius, the poloidal field ends up drifting equatorward.
This figure also shows the appearance of smaller scale structures in the radial field at the surface. In particular, we note a small equatorward branch (about 80 times less intense than the value near the pole) in a very narrow band around the equator (between $-20 \degr$ and $20 \degr$ in latitude) which is of opposite polarity than that of the present cycle. This branch is the remnant of a small amount of field of the preceeding cycle which was driven back up to the surface by the upper cell which creates an upflow at mid-depth near the equator (see the radial velocity profile on Fig. $\ref{fig:courant2ray}$).

\begin{figure}[htbp]
   \centering
\includegraphics[width=9.0cm]{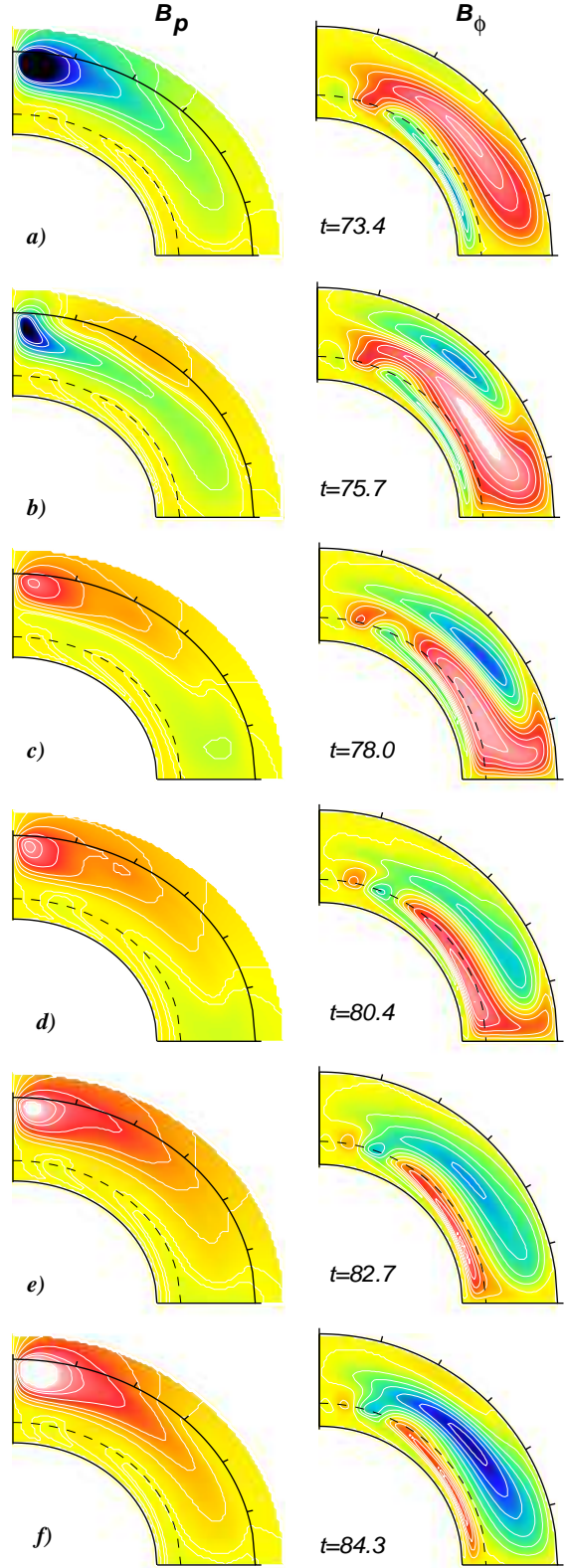}
      \caption{ Case 1b: temporal evolution of the poloidal potential (left panel) and the toroidal field contours (right panel) in a meridional plane for case 1 during half a magnetic cycle. The blue contours indicate an anticlockwise orientation for $B_{p}$ and a negative orientation for $B_{\phi}$ (i.e the field is directed towards the reader in this case) and the red ones clockwise orientation for $B_{p}$ and a positive orientation for $B_{\phi}$ (directed away from the reader).}
       \label{figure_evol2ray}
   \end{figure}

The dependance of the period of this model on variation of the magnetic Reynolds number and thus of the velocity amplitude is very strong in comparison to the unicellular model. In this case, we have the following dependance for the period: $T\propto v_{0}^{-0.93}$.  
The strong dependance of the cycle period on the MC amplitude suggests that it would be easy to recover a solar period of about 22 years, only by increasing the amplitude of the meridional flow, keeping the other parameters constant. 
So the maximum latitudinal velocity $v_{0}$ needed to get a 22-yr cycle period keeping  $s_{0}=20 \, \rm cm.s^{-1}, \eta_{t}=5\times10^{10} \, \rm cm^2.s^{-1}$ would be about $2500 \, \rm cm.s^{-1}$. However, if we only increase the MC amplitude, we lose the antisymmetry of the toroidal field with respect to the equator observed in the Sun (see Sect. \ref{parity}). Thus, to keep the correct dipolar parity for this 2 radial cell model, we need to increase both the MC amplitude and the magnetic diffusivity, hence, we get a solar-like parity 22-yr model with the following parameters: $s_{0}=20 \, \rm cm.s^{-1}, \eta_{\rm t}=1.49\times10^{11} \, \rm cm^2.s^{-1}$ and $1916 \, \rm cm.s^{-1}$.
In Fig. $\ref{figure_dephas2ray}$, we represent the butterfly diagram for this 22-yr cycle case which also exhibits the phase relationship between the poloidal and toroidal fields. We maintain the equatorward branch for the toroidal field between $0$ and $30 \degr$. Moreover, the radial field evolution is smoothened by the increased diffusivity and we thus see much fewer small structures in the lower panel of Fig. $\ref{figure_dephas2ray}$. We nevertheless note that we keep very strong values for the polar field at the surface, which was also the case for the reference model.

Adding a new cell in radius modifies the magnetic advective path and thus the link between the two source regions (the surface and the base of the convection zone). As a consequence, the time-delay between the reversal of the polar field at the surface and the maximum of toroidal field at the base of the CZ is also modified. Indeed, the phase shift between the 2 components of the magnetic field is here about $\pi/3$: we observe that as the poloidal field reverses at the pole, the toroidal field has not yet reached its maximum, thus lagging the poloidal field.

In Fig. $\ref{figure_evol2ray}$, we show the field evolution in the meridional plane of the 22-year cycle model. We see that the field configuration tends to follow the complex nature of the MC. Indeed, we clearly see that we are in the advection-dominated regime as the major field concentration areas follow the meridional flow streamlines. Figure $\ref{fig:courant2ray}$ indicates that we have an upper cell with a poleward flow very concentrated in a thin layer near the surface which we recover on the magnetic patterns especially on panels $c)$ and $d)$ where a small part of the toroidal field is being advected towards the pole near the surface. Between $0.73$ and $0.94 R_{\sun}$ (i.e in more than $60\%$ of the CZ), where the flow is equatorward, most toroidal field of a chosen polarity is advected toward the equator and amplified by the latitudinal shear of the poloidal field. As it reaches the equator, it splits in two parts as we see on panel $d)$, one being redirected towards the surface where it will be driven in the direction of the pole by the top of the upper meridional cell and the other part, containing most of the toroidal field, being advected towards the base of the CZ. As the toroidal field reaches the base of the CZ, the poleward flow advects the field towards the pole [panels $e)$ and $f)$] where it is amplified by the radial shear of poloidal field which at the same time makes the opposite polarity of the preceeding cycle decay away.
We can also see that the poleward branch of the toroidal field of one cycle (cycle $n$) is ``pushed'' towards the pole both by the field of cycle $n-1$ and cycle $n-2$, all present at the base of the CZ at the same period of time. Panels $d)$, $e)$ and $f)$ show that the fields of cycle $n$ and $n-2$, of the same polarity, even reconnect with each other at a given time and stay connected during almost half a magnetic cycle, before being split back by the diverging cells of meridional flow. 

We thus see that having many cells in radius impacts significantly BL models. We now turn to studying the coupled effect of having two cells in radius and latitude.

\subsubsection{Case 2: 2 cells in radius and 2 cells in latitude per hemisphere}

\begin{figure}[h]
\begin{center}
\includegraphics[width=8cm]{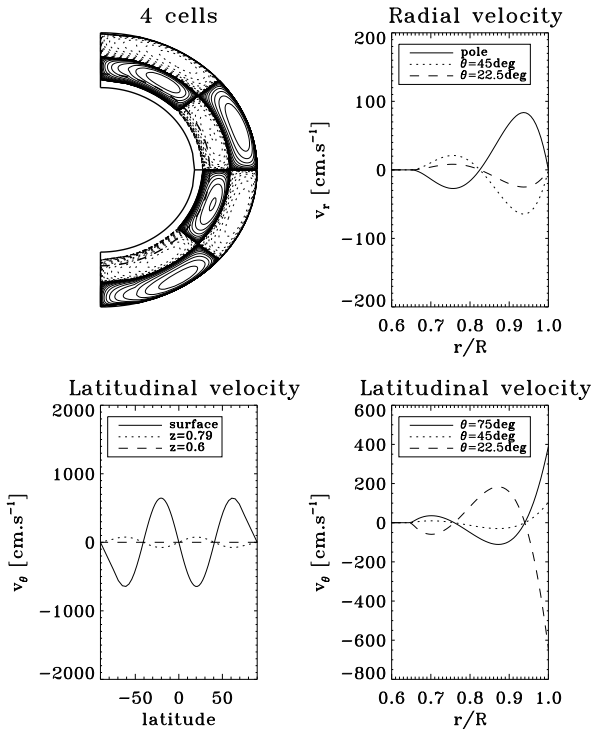}
\end{center}
\caption{ Stream function and components of the meridional flow multiplied by $v_{0}=643 \, \rm cm.s^{-1}$ for case 2.}
\label{fig:courant4cell}
\end{figure}

In this model, the stream function $\psi$ is a product of polynomials of higher order (degree 5 in $z$ and $x$) that we have to multiply by a function of $x$ which enables us to choose the relative velocity amplitude in each cell. We set the 2 top cells to the same maximum value which implies that $\psi$ has the following expression:

\begin{eqnarray}
\psi(x,z)&=&K_{2}z(z-0.65)(500(z-0.825)^3-500\times(0.175)^2   \nonumber \\
&\times&(z-0.825))\times (7x^5-10x^3+3x)\times(1-x^2)^{(1/3.75)} \nonumber
\end{eqnarray}
if $z>0.65$ and 0 otherwise (see fig $\ref{fig:courant4cell}$).

\begin{figure}[h]
   \centering
\includegraphics[width=9cm]{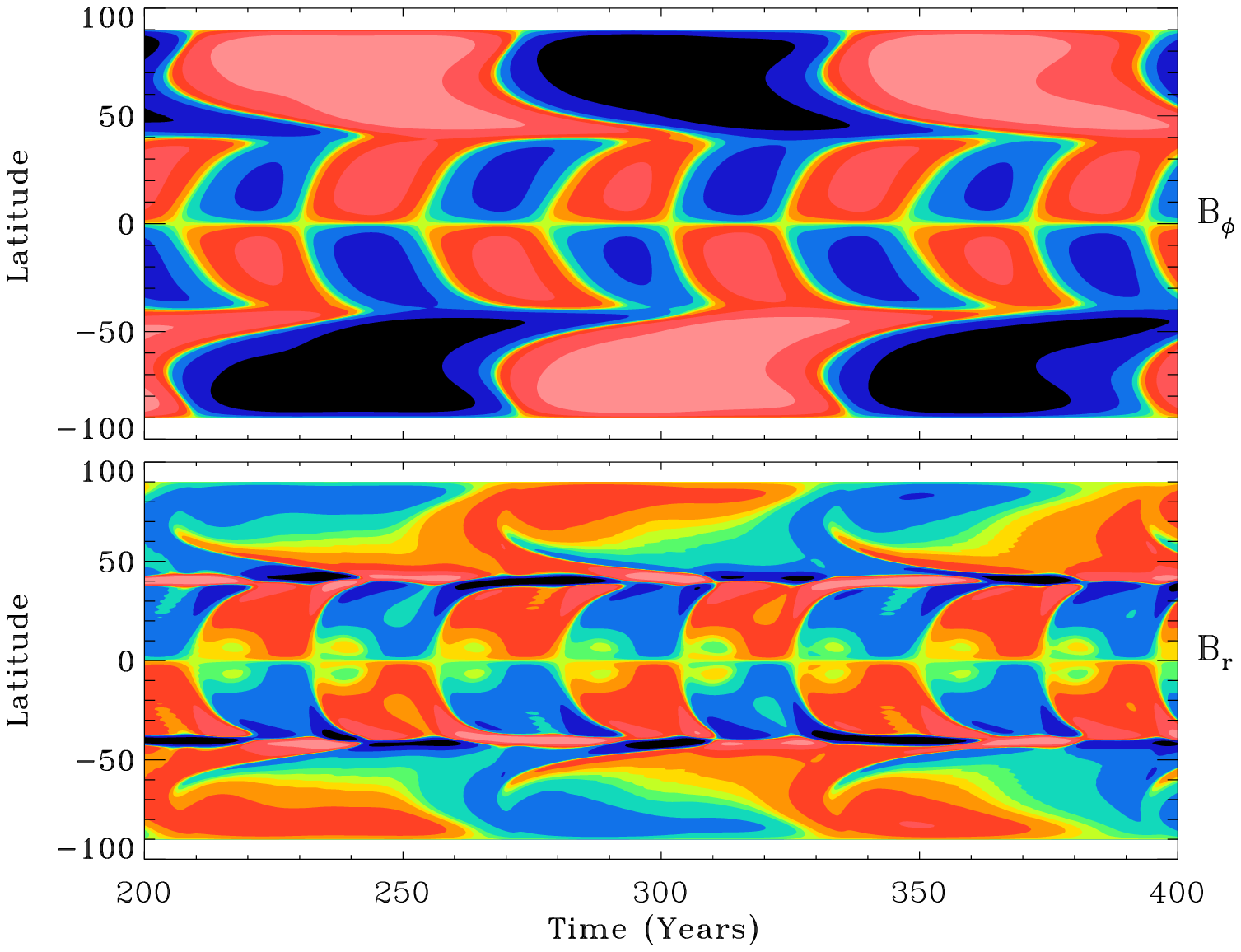}
      \caption{ Case 2a: butterfly diagram (time-latitude cut at $r=\rm cst$) of case 2 with $v_{0}=643 \, \rm cm.s^{-1}$. The format is the same as Fig. $\ref{figure_butterflyA}$.}
       \label{figure_butterfly4cel}
   \end{figure}

\begin{figure}[h]
   \centering
\includegraphics[width=9cm]{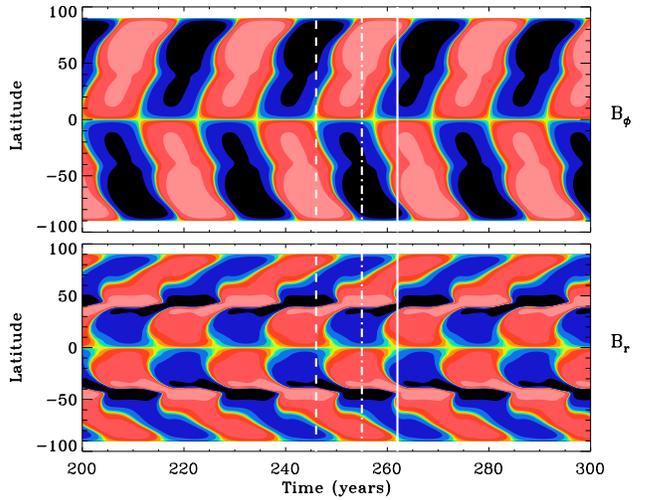}
      \caption{Case 2b: Butterfly diagram (time-latitude cut at $r=\rm cst$) and field phase relation of case 2 with a period close to the solar period.}
       \label{figure_dephas4cel}
   \end{figure}

For the 4 cells model, the butterfly diagram corresponding to the exact same parameters  $\eta_{\rm t}$, $s_{0}$ and $v_{0}$ as the unicellular model is shown in Fig. $\ref{figure_butterfly4cel}$. The most obvious property of this model is that it can sustain two magnetic cycles, one near the equator and the other at high latitudes with significantly different periods. In both equator and polar branches, the cycle period is strongly increased, up to $44.7$ years near the equator and reaching a period of $124$ years at high latitudes. This behaviour is due to the presence of 2 cells in radius which significantly increases the period probably because, as in case 1, the magnetic flux is not transported from the surface to the interface as fast as it was in the unicellular model because a return flow is present at mid depth. However, adding cells in latitude decreases the time for the fluid to travel along the ``conveyor belt'' for the low-latitude region, hence the flux is transported faster from the surface to the base of the convection zone and thus the regeneration of each component of the magnetic field is faster, as shown by Dikpati et al. (\cite{Dikpati2}). Here, by having cells both in radius and in latitude, we get a cycle faster than the two radial cells case but still slower than the unicellular case. Consequently, the influence of having several radial cells seems to be much stronger than that of adding cells in latitude.
We can see on this butterfly diagram that the patterns are quite complex in the time-latitude plane. We clearly see on the toroidal field at the base of the CZ and on the poloidal field at the surface the imprint of the 2 counter-cells at $45 \degr $. Once again the magnetic field behaviour strongly depends on the direction of the flow. Indeed, at the base of the CZ, the toroidal field at high latitudes (where the flow is equatorward) is drifting from $60 \degr$ to $45 \degr$ where it encounters the counter latitudinal cell. At low latitudes (where the flow is poleward), the field is advected polewards from the equator to the zone of vanishing $v_{\theta}$ (i.e. around $45 \degr$).
For the radial field, the patterns are also very intricate. We again find the small equatorward branch as in case 1. We also clearly see the generation of new poloidal field structures just above $50 \degr$ embedded in the structure of opposite polarity. It is then moving down to $45 \degr$, meeting a branch of the same polarity coming from the preceeding cycle and from the equator. The main part of the radial field at the surface is then advected towards the pole, creating a significant polar branch in the high-latitudes part. One interesting feature of this model is that the intensity of the polar field is diminished by a factor 10 compared to the unicell model, which could be in better agreement with the solar observations.   

 We tested the sensitivity of the multicellular model's characteristics to variations of the physical parameters. Using a least square fit to get the exponents of each parameter, we note that the dependance of the cycle period at low latitudes on $s_{0}, v_{0}$ and $\eta_{\rm t}$ is as follows:

$$
T \propto s_{0}^{0.05} v_{0}^{-0.35} \eta_{\rm t}^{-0.4}
$$

The dependance of the period of this model to a variation of the velocity amplitude is reduced in comparison to the previous models.
The cycle period will thus be less disturbed by temporal fluctuations of the MC amplitude, which is a very attractive feature of this model. 
For any value of $s_{0}$, we expect that an increase in the intensity of the meridional flow enables the field to travel faster along the `conveyor belt' so that both poloidal and toroidal fields head faster towards their reversals. This is why we have a negative dependance of the period on $v_{0}$.

\bigskip

As soon as the Reynolds number becomes too high ($R_{\rm e}$ above $800$), i.e. when the strength of the meridional flow is increased, $\eta_{\rm t}$ remaining constant, a strong polar branch with a longer period is appearing. This property is due to the advection by meridional flow of the magnetic field in the whole convection zone dominated by its poleward component (see below). When the field reaches the base of the CZ, the strong toroidal structure is thus concentrated near the pole, trapped in the slowly moving meridional cell at high latitudes. As we do not get this feature of two coexisting branches of different periodicity in the Sun, we seek to recover a unique cycle period, taking into account the dependance on parameters. We have seen that it is not sufficient to act on the strength of the meridional circulation to recover the 22-yr cycle period, but as the least square fit shows, turbulent diffusivity plays a crucial role in this multicellular model.

\bigskip 

The strong negative dependance of the cycle period on the turbulent magnetic diffusivity is characteristic of the fact that we have multiple cells of meridional flow in each hemisphere. 
The magnetic field follows the configuration of the meridional flow as long as it is advected by the circulation but the magnetic diffusivity enables the field to cross the strong velocity gradients present at the borders of each cell. Magnetic diffusivity thus provides the field a way to short-circuit the complex ``advection path'' of this model and to allow for a faster link between one meridional cell and another. As a consequence, if $\eta_{\rm t}$ is too low, the poloidal field created at the surface will cross the strong velocity gradients that we have at the borders between the various cells in a much longer time and the classical mechanism of regeneration of toroidal field from this existing poloidal field will be less efficient. 
Consequently, to obtain a 22-yr cycle period for this model, we have to increase the magnetic diffusivity (without of course losing the advection-dominated regime) to $1.5 \times 10^{11} \, \rm cm^2.s^{-1}$ (so that the magnetic field is enabled to get from one circulation cell to another quickly enough) and in agreement with the least-square fit which indicates a negligeable dependance of the period on $s_{0}$ and a negative dependance on the velocity amplitude, we keep the source term intensity to $s_{0}=20 \, \rm cm.s^{-1}$, and slightly increase $v_{0}$ to $1071 \, \rm cm.s^{-1}$. Figure $\ref{figure_dephas4cel}$ shows the butterfly diagram and the phase relationship between the poloidal and toroidal fields for this 22-yr period case. 
The first positive result of this model is that no second cycle period appears, we get a unique cycle due to the dynamo action in the whole CZ.
Here again, as the link between the 2 source regions of poloidal and toroidal fields is complex, the field relation is not that which we observe in the Sun and it is even more perturbed than in case 1. Indeed, the poloidal field here reverses approximately where the toroidal field of the opposite polarity reverses, meaning that we have a phase shift of about $\pi$ between the 2 components of the field, with the toroidal field leading the poloidal field. We note that in this case, unlike case $1$, the shape of the butterfly diagram is slightly different from that using the parameters of the single cell model. Indeed, increasing the magnetic diffusivity caused the sharp magnetic structures to be smoothed, especially for the radial field in which the small equatorward branch of opposite polarity is not visible anymore and where the branch at $50 \degr$ drifting to $45 \degr$ is thicker and smoother. 

\begin{figure}[htbp]
   \centering
\includegraphics[width=9.cm]{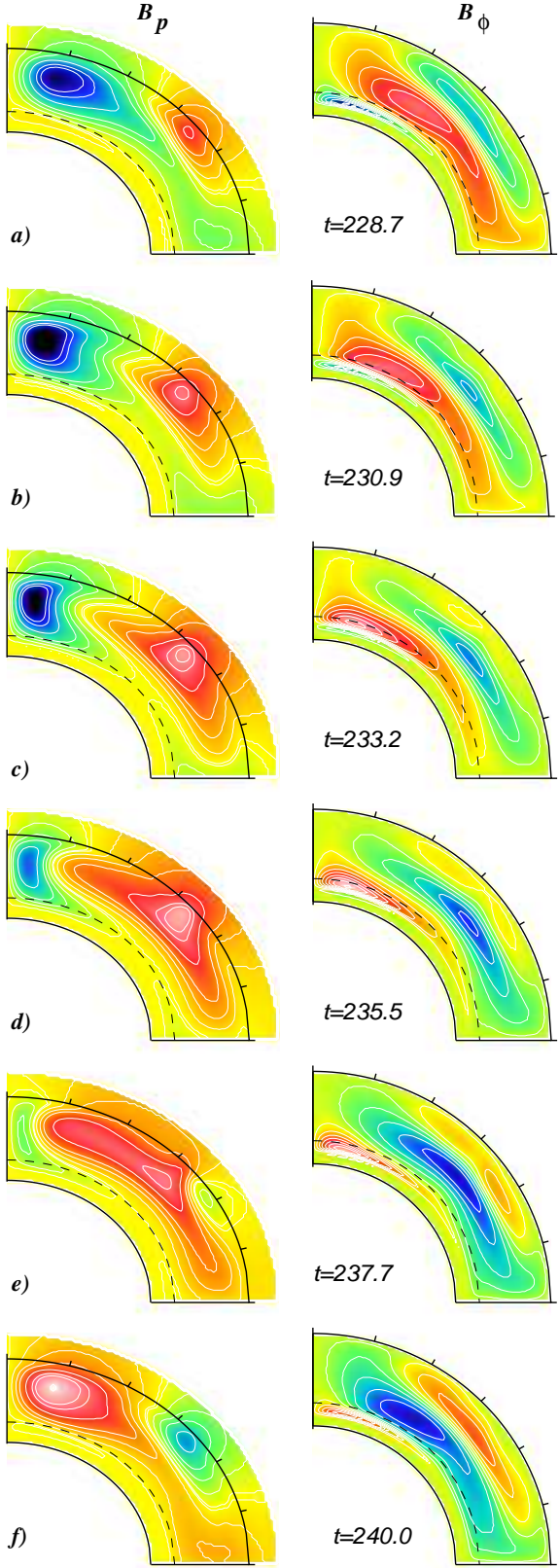}
      \caption{Case 2b: temporal evolution of the poloidal field lines with its potential extrapolation (left panel) and the toroidal field contours (left panel) in a meridional plane for case 2 for half a magnetic cycle. The format is identical to Fig. $\ref{figure_evol2ray}$.}
       \label{figure_evol4cel}
   \end{figure}

Figure $\ref{figure_evol4cel}$ shows the temporal evolution of the magnetic field in the meridional plane for the same case 2b. The first appearance of a new poloidal field occurs at a latitude of about $30 \degr$, created by the BL source term. The creation of poloidal field at such latitudes is a direct consequence of the non-locality of our source term whose latitudinal dependance is linked to the latitudinal dependance of the toroidal field existing at the base of the CZ at the same time. This poloidal field structure is then both slowly dragged towards the $45 \degr$ latitude by the MC and amplified by the source term which is still active in this region. The poloidal field is here latitudinally sheared to create and amplify toroidal field [see panels $a)$ and $b)$]. After the reconnection, MC has dragged the field deeper down to regions of both poleward and equatorward flow so the field is located at the particular convergence point between the 4 cells. This inward advection due to the meridional flow configuration at $45 \degr$ is also acting on the toroidal field to drive it down to the middle of the CZ and should not be mistaken with diffusive effect. It is really the strongly negative radial velocity at this latitude (see panel 2 of Fig. \ref{fig:courant4cell}) which advects the field inward to the point of convergence of the 4 cells. We see on panels $c)$, $d)$ and $e)$ that even if some of the field is advected towards the equator, the north branch is dominant and thus most of the field goes up to the pole, while the older field of opposite polarity is being advected to the base of the CZ. This poloidal field of opposite polarity (the negative field on panels $d)$, $e)$ and $f)$) is then heading back to the $45 \degr$ region where it meets another poloidal structure of the same polarity.
If we look at the toroidal field, we see it is created and amplified in the whole convection by the latitudinal shear of the differential rotation. At the same time, $B_{\phi}$ is being advected by the flow, first inward to the middle of the CZ and then mainly to the polar regions. It is finally dragged down to the base of the CZ where the field of the preceeding cycle is cancelled by the creation of a new field of opposite polarity. 
Unlike case 1, the suppression of the field of the preceeding cycle seems thus to be mainly due to the advection of the present $B_{\phi}$ (which was created before by the shear in latitude of the poloidal field) than to the radial shear of $B_{p}$ in the tachocline. 
In this case, only cycle $n$ and $n-1$ seem to really interact since the toroidal field of cycle $n-2$ has already completely vanished at the bottom of the CZ when $B_{\phi}$ of cycle $n$ is being created in the upper part of the CZ.

This 4 cells model is thus very intriguing for many important solar dynamo properties. The butterfly diagram as well as the field lines evolution during a cycle become very complex, the field phase relationship is not corresponding anymore to the solar observations but the dependance on the MC amplitude is reduced and the strength of the polar field is decreased, which constitute attractive characteristics of the model.

\section{Parity selection in multiple meridional cells dynamo models}
\label{parity}

 As we said before, Dikpati $\&$ Gilman in 2001 showed that with a set of parameters they found appropriate to give a solar-like solution, their pure BL flux transport model had difficulties reproducing the persistent antisymmetry of the toroidal field we observe in the Sun.

Several solutions were proposed to solve this problem, Dikpati $\&$ Gilman (\cite{Dikpati5}) as well as Bonanno et al. (\cite{Bonanno1}) managed to get rid of this field parity drift by imposing an $\alpha$-effect at the base of the CZ, thus imposing two spatially separated source terms for the poloidal field. Another solution was proposed by Chatterjee et al. (\cite{Chatterjee}): they keep the regular surface source term of BL type but they impose a small diffusivity ($2.2 \times 10^8 \, \rm cm^2.s^{-1}$) in the overshoot layer to prevent the toroidal field from diffusing across the equator and a very large diffusivity ($2.4 \times 10^{12} \, \rm cm^2.s^{-1}$) for the poloidal field in the convective zone to allow diffusive coupling of the poloidal field between the two hemispheres.

We now seek to characterise the influence of multicellular flows on parity selection and we consider its sensitivity to variations of different parameters. The results are summarized in Tables $\ref{table_5e10}$ and $\ref{table_8e10}$.

Computing the critical dynamo numbers (the threshold value of $C_{\rm s}$ for which the magnetic energy begins to grow), starting from a dipolar configuration and then from a quadrupolar one, enables us to test the influence of the MC amplitude and of the diffusivity on the parity selection in our various cases. 
 In a relatively low range of MC amplitude ($v_{0}<1000 \, \rm cm.s^{-1}$) and at a magnetic diffusivity of $5.10^{10} \, \rm cm^2.s^{-1}$, dipolar solutions are easily excited in the unicellular case. On the contrary, for the multicellular models, the symmetric parity is already appearing at lower values of $v_{0}$. Indeed, the magnetic field has switched to a quadrupolar parity at $v_{0}=785 \, \rm cm.s^{-1}$ for the two radial cell model and the difference between $C_{\rm s}^{\rm cr}(A)$ and $C_{\rm s}^{\rm cr}(S)$ is already very small at $v_{0}=643 \, \rm cm.s^{-1}$ ($0.84$ compared to $0.85$). The 4-cell model always favours the quadrupolar configuration, even for low velocities.

\begin{table}[h]
\begin{center}
\caption{Critical values of $C_{\rm s}$ starting from a dipole (A) or a quadrupole (S) for various values of $v_{0}$ and at $\eta_{\rm t}=5.10^{10} \, \rm cm^2.s^{-1}$ for the 3 configurations of meridional circulation. The favoured symmetry (the smallest values of $C_{\rm s}$) is indicated in bold characters. In the first line the $C_{\rm s}^{\rm crit}$ for the reference case, case 1a and case 2a are shown.}
\begin{tabular}{c|cc|cc|cc} \hline 
 & \multicolumn{2}{|c} {Single Cell} & \multicolumn{2}{|c} {Two radial cells} &\multicolumn{2}{|c} {4 cells} \\ \hline
 $v_{0}$ & $C_{\rm s}^{\rm Cr}(A)$ & $C_{\rm s}^{\rm Cr}(S)$ & $C_{\rm s}^{\rm Cr}(A)$ & $C_{\rm s}^{\rm Cr}(S)$ & $C_{\rm s}^{\rm Cr}(A)$ & $C_{\rm s}^{\rm Cr}(S)$ \\ \hline
$643$ & {\bf 1.88} & 1.92 & {\bf 0.84} & 0.85 & 1.16 & {\bf 1.01} \\ \hline
$785$ & {\bf 2.34} & 2.37 & 1.05 & {\bf 1.04} &  1.19 & {\bf 1.05} \\ \hline
$1000$ & 3.07 & {\bf 3.05} & 1.40 & {\bf 1.35} & 1.55 & {\bf 1.38} \\ \hline
$1500$ & 4.51 & {\bf 4.49} & 2.50 & {\bf 2.20} & 3.00 & {\bf 2.73} \\ \hline
\end{tabular}
\label{table_5e10}
\end{center}
\end{table}

 To obtain a smaller $C_{\rm s}^{\rm crit}$ for the dipolar mode in the multiple cell models, we need to increase the magnetic diffusivity. Indeed, for all cases, increasing the diffusivity widens the range for $v_{0}$ in which we stay in the dipolar configuration. In case 1, at $\eta_{\rm t}=5.10^{10} \, \rm cm^2.s^{-1}$ and $v_{0}=1000 \, \rm cm.s^{-1}$, $C_{\rm s}^{\rm crit}=1.4$ for the dipole and $C_{\rm s}^{\rm crit}=1.35$ for the quadrupole, which explains the drift of parity we observe in this case in Table  $\ref{table_5e10}$. When we increase the magnetic diffusivity up to $\eta_{\rm t}=8.10^{10} \, \rm cm^2.s^{-1}$, the dipole becomes the most easily excited solution with $C_{\rm s}^{\rm crit}(\rm dipole)=1.29$ and $C_{s}^{\rm crit}(\rm quadrupole)=1.3$. In the same way, the systematic parity switching in the 4-cell model disappear when we go from $\eta_{\rm t}=5.10^{10} \, \rm cm^2.s^{-1}$ to $\eta_{\rm t}=8.10^{10} \, \rm cm^2.s^{-1}$ and for example at $v_{0}=785 \, \rm cm.s^{-1}$, the solar symmetry is favoured. We thus confirm the work of Chatterjee et al. (\cite{Chatterjee}) which shows that increasing the diffusivity in the convection zone, thus allowing diffusive coupling of the poloidal field between the two hemispheres improves the parity conservation.

\begin{table}[h]
\begin{center}
\caption{Critical values of $C_{\rm s}$ starting from a dipole (A) or a quadrupole (S) for various values of $v_{0}$ and at $\eta_{\rm t}=8.10^{10} \, \rm cm^2.s^{-1}$ for the 3 configurations of meridional circulation. The favoured symmetry (the smallest values of $C_{\rm s}$) is indicated in bold characters.}
\begin{tabular}{c|cc|cc|cc} \hline 
 & \multicolumn{2}{|c} {Single Cell} & \multicolumn{2}{|c} {Two radial cells} &\multicolumn{2}{|c} {4 cells} \\ \hline
$v_{0}$ & $C_{\rm s}^{\rm Cr}(A)$ & $C_{\rm s}^{\rm Cr}(S)$ & $C_{\rm s}^{\rm Cr}(A)$ & $C_{\rm s}^{\rm Cr}(S)$ & $C_{\rm s}^{\rm Cr}(A)$ & $C_{\rm s}^{\rm Cr}(S)$ \\ \hline
$643$ & {\bf 2.72} & 2.81 & {\bf 0.79} & 0.85 & {\bf 2.10} &  2.14 \\ \hline
$785$ & {\bf 2.79} & 2.95 & {\bf 0.99} & 1.03 &  {\bf 2.11} &  2.15 \\ \hline
$1000$ & {\bf 2.89} & 3.08 & {\bf 1.29} & 1.30 & 1.80 & {\bf 1.50} \\ \hline
$1500$ & {\bf 4.41} & 4.43 & 1.96 & {\bf 1.94} & 2.12 & {\bf 1.92} \\ \hline
\end{tabular}
\label{table_8e10}
\end{center}
\end{table}

\begin{figure}[h]
   \centering
   \includegraphics[width=9cm]{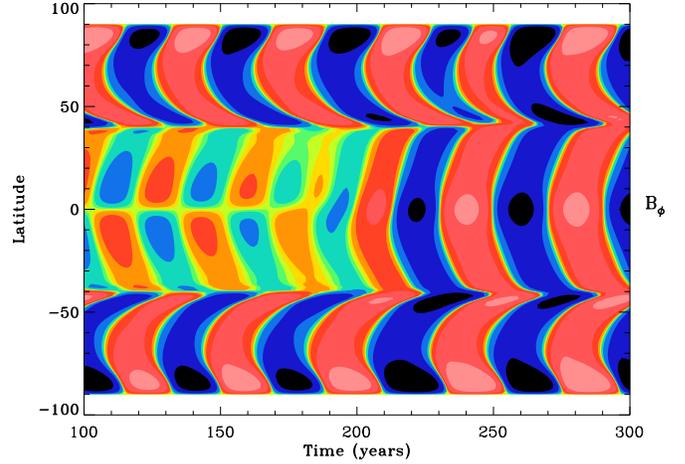}
      \caption{Zoom on the epoch of parity drift of the toroidal field for the 4-cell model with $u_{0}=2000 \, \rm cm.s^{-1}$, $s_{0}=20 \, \rm cm.s^{-1}$ and $\eta_{\rm t}=5.10^{10} \, \rm cm ^2.s^{-1}$.}
       \label{fig_changeparity}
   \end{figure}

However, as soon as we increase the amplitude of the velocity field via an increase of the MC amplitude, we recover the parity drift from a dipolar to a quadrupolar configuration, the quadrupole becomes the easiest solution to excite. Figure $\ref{fig_changeparity}$ shows a typical representation of a parity shift in the case of 4 cells of meridional circulation per hemisphere: the toroidal field at the base of the CZ is switching from an antisymmetric configuration with respect to the equator to a symmetric one.  As Dikpati $\&$ Gilman (\cite{Dikpati5}) showed, it is the connection at the equator of the sufficiently strong poloidal fields of each hemisphere that enables the cancellation of $B_{r}$ and the creation of an antisymmetric $B_{\theta}$. The shear of this antisymmetric $B_{\theta}$ by differential rotation is then responsible for the creation of antisymmetric toroidal field which we observe in the Sun. Thus it is very likely that increasing the velocity amplitude makes the magnetic field travel faster along the conveyor belt and prevents the poloidal field from staying in the same location enough time to connect with its counterparts in the opposite hemisphere. As a consequence, models with faster flows shift to quadrupolar solution since they prevent these reconnection phenomena. 

This particular property of parity selection explains why it is not sufficient to act on the MC amplitude in case 1 to recover a satisfying model with a 22-yr cycle. Indeed, increasing the velocity causes the magnetic field to become symmetric with respect to the equator. On the contrary, we saw that increasing the diffusivity tends to favour dipolar symmetry. As a consequence, cases 1b and 2b, which have relatively strong values of the diffusivity are able to reproduce the field antisymmetry we observe in the Sun, even with significant meridional flow amplitudes. For case 1b, the values of the $C_{\rm s}^{\rm crit}$ are $C_{\rm s}^{\rm Cr}(A)=2.47$ and $C_{\rm s}^{\rm Cr}(S)=2.48$ and for case 2b, we get $C_{\rm s}^{\rm Cr}(A)=3.36$ and $C_{\rm s}^{\rm Cr}(S)=3.39$. These cases are thus able to sustain a 22-yr cycle period without drifting from a dipolar to a quadrupolar configuration.

\section{Conclusion and perspectives}

In this paper, we have discussed 2D BL flux transport type solar dynamo models with various profiles of meridional flows.

We have first tested the influence of introducing the parameters giving a solar-like solution in the reference unicellular case in the muticellular models. These cases, denoted 1a and 2a, show that the presence of a multicellular circulation has a strong perturbing impact on the behaviour of solar dynamo models.
Adding cells in radius (case 1a) leads to a complex advective path and thus causes the cycle period to be more than tripled compared to the reference case (the cycle is here of 84.6 years instead of 22 years in the reference model) and in this case, a strong poleward branch appears on the butterfly diagram, due to the poleward advection by the MC at the base of the CZ. The radial field at the surface seems to show very fine and small structures during its whole cyclic evolution. In this model, we notice that the cycle period is moreover strongly linked to the amplitude of the meridional flow, which indicates that the cycle will be significantly sensitive to the observed fluctuations in the MC amplitude. In the 4-cell model (case 2a), the most obvious property is that we seem to get two magnetic cycles, with different periods, both longer than in the reference case (44.7 years near the equator and 124 years near the poles). Moreover, the phase relationship between the poloidal and toroidal parts of the field does not match the solar observations anymore. However, unlike case 1a, the dependance on the amplitude of the MC is reduced, which could make the model and thus the cyclic activity more robust and less sensitive to temporal fluctuations observed in the Sun. 

The set of parameters were in these cases clearly not adapted to recover a 22-yr cycle period, we thus modified the appropriate parameters to get a solar-like solution. Relying on the least-square fits, cases 1b and 2b were thus computed with higher diffusivities and higher MC amplitudes and a 22-yr cycle was indeed recovered. It should be noted that we stay in these cases in realistic values for both magnetic diffusivity and MC amplitude and that our models are still all dominated by advection, as shown on the evolutions of the field lines in the meridional plane. For these cases, we note that the butterfly diagrams are smoothed out probably thanks to the increase of magnetic diffusivity, the small structures visible on the radial field on cases 1a and 2a vanish, we thus seem to obtain a globally less perturbed behaviour for these models. In case 2b, we moreover decrease the intensity of the polar surface field, which is in better agreement with observations. But we still obtain properties such as the phase relationship between the poloidal and toroidal parts of the field that are not that which we observe in the Sun. 
Since parity selection is one of the major concerns in BL flux transport dynamo models, we focus in Sect. \ref{parity} on the parity issue in these models. We show that adding cells both in radius and in latitude seems to favour the quadrupolar parity, which we do not observe in the Sun. However, if the magnetic diffusivity is sufficiently high, we get diffusive coupling of the poloidal field across the equator and thus the dipolar parity conservation is improved. On the contrary, if the MC amplitude is increased, the major trend of all cases is to switch from a dipolar to a quadrupolar magnetic field configuration. We can nevertheless recover cases with complex meridional flow with a 22-yr cycle period and a favoured dipolar parity, staying in realistic values of both MC amplitude and magnetic diffusivity, this is the case for models 1b and 2b.

As far as the meridional flow amplitude is concerned, we shall note that our relatively small velocities at the solar surface are related to the small stratification we have in our models (our density is proportional to $1/r$) which implies a small velocity contrast between the surface and the bottom of the convection zone. Indeed, the density profile used by Dikpati \& Charbonneau (\cite{Dikpati1}) was proportional to $\sqrt{R_{\sun}/r-1}$ which has a strong variation in radius especially near the surface. Thus, the most important velocity amplitude (that at the base of the CZ, which advects the strong toroidal field created by differential rotation) can be identical with very different velocities at the surface in these two models using different density profiles. We also note that since there was a factor 2 between $v_{0}$ and ${\rm max}(v_{\theta})$ in the work of Dikpati \& Charbonneau (\cite{Dikpati1}), a direct comparison with our work (where $v_{0}={\rm max}(v_{\theta})$) is not straightforward. 

Even if cases 1b and 2b seem to allow a 22-yr cycle combined with persistent solar-like dipolar parity and a smooth cyclic butterfly diagram, we show that introducing a complex MC in our models has a strong perturbing impact, for example on the phase relationship between the poloidal and toroidal parts of the magnetic field which does not correspond to the solar observations. We are thus heading to the hypothesis that the BL mechanism may not be the only source of poloidal field in the solar dynamo cycle, especially if the Sun happens to show a persistent multicellular meridional flow which seems to be quite destructive for several solar cycle features in the pure BL flux-transport framework. Of course we now need to check the influence of a less monolithic meridional flow structure, with extra cells more concentrated in a particular area of the CZ and varying in time since the position and strength of each meridional cell seem to influence quite significantly the global properties of the solar dynamo.

It is thus now a real challenge for local helioseismology to probe the Sun deeper to give better constraints on the meridional flow in the convection zone.

\begin{acknowledgements}

We wish to thank Paul Charbonneau (Universit\'e de Montreal) for the original version of the numerical code. We also gratefully acknowledge useful discussions with Mausumi Dikpati (HAO/NCAR) and thank her for giving us precious advice.

\end{acknowledgements}

\newpage


\begin{appendix}

\section{Numerical approach}

To solve the equations, we use a code adapted by P. Charbonneau and T. Emonet from Finite Element Analysis by D.S.Burnett (\cite{Burnett}). This code enables us to solve a general partial differential equation (PDE) using a finite element method in space and a third order scheme in time. We adapted it to problems such as $\alpha^2-\Omega$, flux transport or multicellular flux transport solar dynamos and we implemented new boundary conditions (radial or potential field at the top) and initial conditions.

\subsection { Spatial method}

The finite element method is a very efficient way to obtain approximate solutions to linear or non linear PDEs in any kind of geometry. Our code STELEM (STellar ELEMents) solves Eqs. $\ref{eqA2}$ and $\ref{eqB2}$ with this method, ie seeking the approximate solutions $\tilde{A_{\phi}}$ and $\tilde{B_{\phi}}$ as linear combinations of trial functions $\psi_{i}$ (to be more specific these trial functions are Lagrange polynomials of degree 1 (linear functions) and serendipity shape functions for second order interpolation (quadratic functions), depending on the complexity of our equations). 

$$
\tilde{A_{\phi}}(r,\theta,t)=\sum_{i=0}^{N} \,\, a_{i}(t)\psi_{i}(r,\theta)
$$

$$
\tilde{B_{\phi}}(r,\theta,t)=\sum_{i=0}^{N} \,\, b_{i}(t)\psi_{i}(r,\theta)
$$

The main steps of the method are the following:

\begin{itemize}
\item Our domain (the annular meridional cut) is divided into smaller regions called {\it elements}. In our case, they are straight-sided quadrilaterals when we use first order Lagrange polynomials, with a node at each corner of the quadrilateral and with a node at each corner and one extra node per side and without any interior node if we use second order interpolation (see Fig. $\ref{figure_mesh}$). 

\begin{figure}[h]
  \centering
\includegraphics[width=3.8cm]{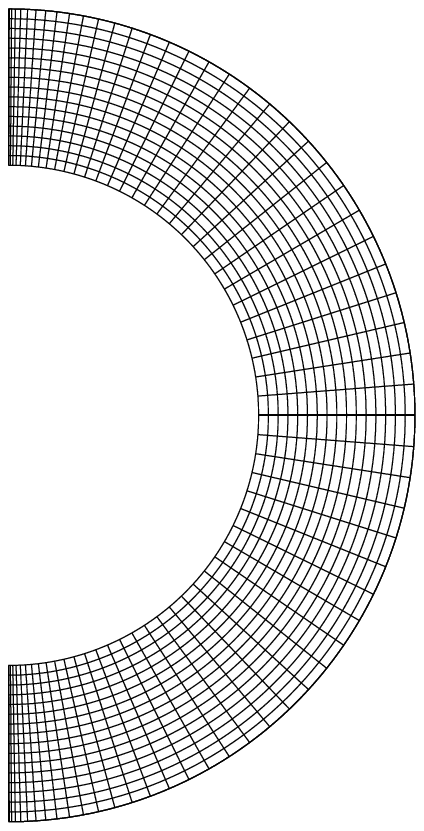}
\includegraphics[width=3.8cm]{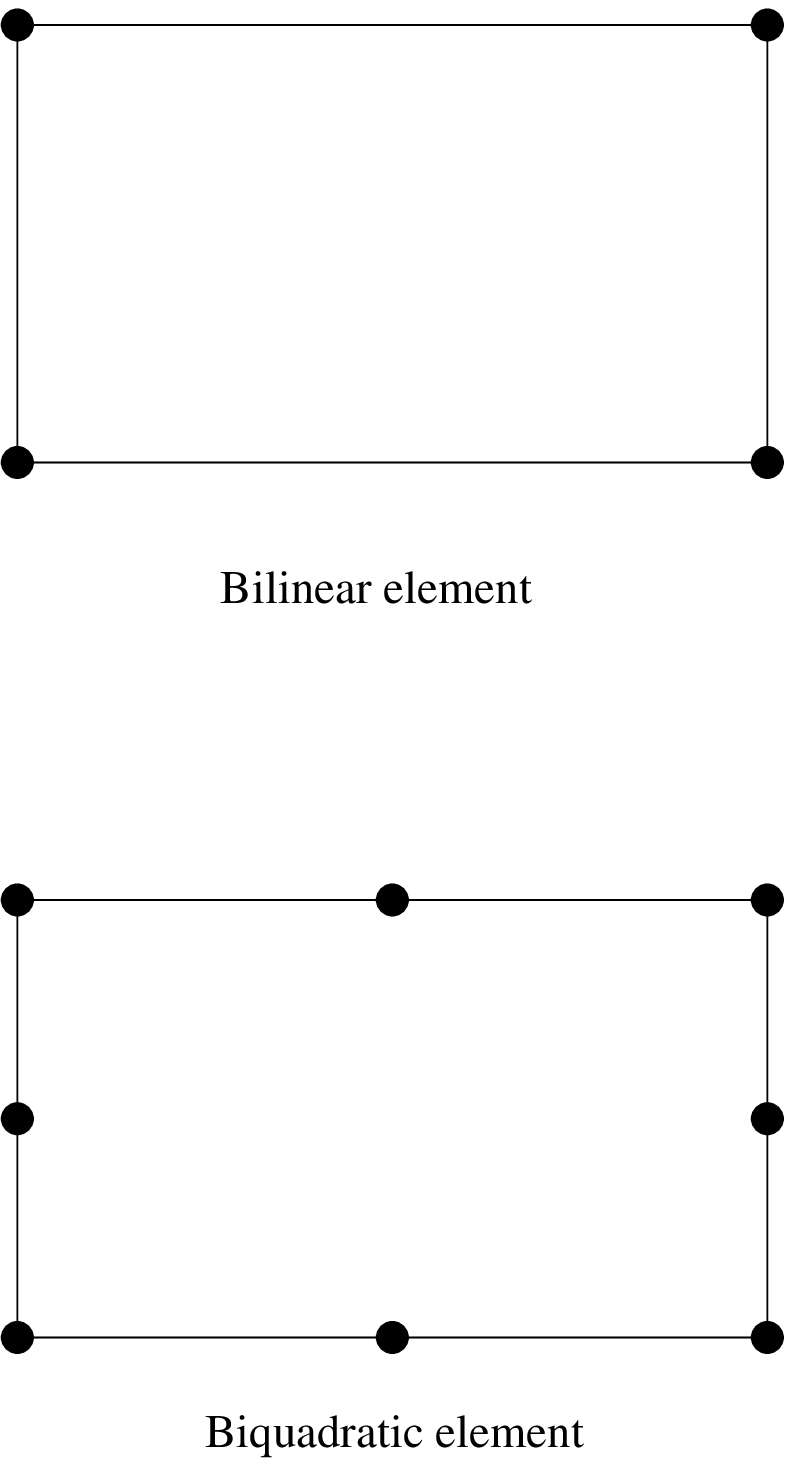}
      \caption{Sketch of the quadrilateral mesh we are using in the meridian plane, uniformly spaced in $r$ and more accurate in the polar regions. As we work with the variables $\cos\theta$,$r$, we get a rectangular grid in the $r$,$\cos\theta$ plane. On the right, we show a zoom on a single rectangular cell with the 4 nodes at each corner in the case of the first order interpolation and with one extra node per side in the case of the second order interpolation. Note that the cells in the quadratic case do not contain any interior node.}
       \label{figure_mesh}
   \end{figure}

\item In each element, the PDEs are transformed into ordinary differential equations (ODE) in time involving the coefficients $a_{i}(t)$ and $b_{i}(t)$ of the linear combinations.
\item The terms in the element equations are numerically evaluated for each element in the mesh. The resulting numbers are assembled into a much larger set of equations called the system equations.
\item The boundary conditions are taken into account. They can be of Dirichlet type (we impose the value of the function at the boundary) or of Neumann type (we impose the normal derivative of the function at the boundary). In particular for the potential extrapolation, we proceed as the following, we were largely inspired by the procedure of Dikpati $\&$ Choudhuri \cite{Dikpati4}.

The top boundary condition is that we have to match smoothly our magnetic field ${\bf B}(r,\theta,t)$ to a field satisfying the free space equation:

$$
{\bf \nab}\times{ \bf B}(r,\theta,t)=0
$$

As we work in spherical axisymmetric geometry, we write that:
$$
{{\bf B}}(r,\theta,t)=\nab\times (A_{\phi}(r,\theta,t) \hat {\bf e}_{\phi})+B_{\phi}(r,\theta,t) \hat {\bf e}_{\phi}
$$

And the equation of free space leads to two equations, one concerning $B_{\phi}(r,\theta,t)$ and the other one concerning $A_{\phi}(r,\theta,t)$, which are:

$$
\frac{\partial (\sin\theta B_{\phi})}{\partial \theta}=\frac{\partial (r B_{\phi})}{\partial r}=0
$$

$$
(\nabla^2-\frac{1}{r^2\sin^2\theta})A_{\phi}=0
$$

As we are dealing with a finite element method, the most convenient and natural procedure is to seek to express these boundary conditions as either Dirichlet or Neumann conditions for $A_{\phi}$ and $B_{\phi}$.

Equation for $B_{\phi}$ very easily leads to the solution $B_{\phi}=C/(r\sin\theta)$, C being a constant. We fix this constant value to $0$ so that our top boundary condition on $B_{\phi}$ is the homogenous Dirichlet condition: $B_{\phi}(R_{\sun},\theta,t)=0$.

We now have to deal with the more difficult condition on $A_{\phi}$. A general solution to this equation can be written in the form 
$$
A_{\phi}(r\geq R_{\sun},\theta,t)=\sum_{n=1}^{+\infty} \frac{a_{n}(t)}{r^{n+1}}P_{n}^1(\cos\theta)
$$
where $P_{n}^1(\cos\theta)$ is the associated Legendre polynomial. We find that truncating the sum at $N_{\theta}/2$, half of the number of grid points in $\theta$ always result in an error in the projection of less than $10^{-3}$. Hence, the coefficients $a_{n}(t)$ are the coefficients of the expansion of $R^{n+1}A_{\phi}(R_{\sun},\theta,t)$ on the associated Legendre polynomials. Thus, the value of $a_{n}(t)$ is calculated by the scalar product of $R^{n+1}A_{\phi}(R_{\sun},\theta,t)$ with $P_{n}^1(\cos\theta)$, divided by the norm of $P_{n}$, leading to:

$$
a_{n}(t)=\frac{R_{\sun}^{n+1}\int_{0}^{\pi}\,\, A_{\phi}(R_{\sun},\theta,t)P_{n}^1(\cos\theta)\sin\theta d\theta}{\int_{0}^{\pi}\,\,[P_{n}^1(\cos\theta)]^2\sin\theta d\theta}
$$

By the variable change $x=\cos(\theta)$ in the upper integral, we are led to calculate the integral on $[-1,1]$ of the product of two smooth functions. Thus we calculate this integral using a Gauss-Chebyshev quadrature formula which uses the weighing functions $1/\sqrt{1-x^2}$. The lower integral is the norm of the associated Legendre polynomials, we know that its value is: $2n(n+1)/(2n+1)$.

From the coefficients $a_{n}(t)$, we can deduce the derivative of $A_{\phi}$ at the solar surface:
$$
\frac{\partial A_{\phi}}{\partial r} |_{r=R_{\sun}}=-\sum_{n=1}^{N_\theta/2}\frac{(n+1)a_{n}(t)}{R_{\sun}^{n+2}}P_{n}^1(\cos\theta)
$$

and from a simple finite difference scheme, we impose a Dirichlet condition on the poloidal potential, calculating the new value of $A_{\phi}$ at the surface, using the points of the layer immediately below the surface. It leads to:

\begin{eqnarray}
A_{\phi}(R_{\sun},\theta,t)&=&A_{\phi}(R_{\sun}-\Delta r,\theta,t)+\Delta r \frac{\partial A}{\partial r} |_{r=R_{\sun}} \\ \nonumber 
&=&A_{\phi}(R_{\sun}-\Delta r,\theta,t)-\Delta r \sum_{n}\frac{(n+1)a_{n}(t)}{R_{\sun}^{n+2}}P_{n}^1(\cos\theta)
\end{eqnarray}

Inside the same time step, we have then to recalculate the coefficients $a_{n}(t)$ with the new value of $A_{\phi}(R_{\sun},\theta,t)$ until we get a sufficiently small difference between two successive values of  $A_{\phi}(R_{\sun},\theta,t)$ in order to make the procedure converge, we usually do not need more than 10 iterations to get a convergence with a relative error of $10^{-3}$.

\item We get a final set of ODEs in time which we solve with a third order scheme we describe below.
\end{itemize}

\subsection { Temporal scheme}

The scheme that we use is adapted from Spalart et al. \cite{Spalart}. We have to solve the following ODE:

$$
\frac{\partial A_{\phi}}{\partial t}={\cal N} (A_{\phi})
$$

$\cal N$ being the non linear operator evaluated in the preceeding step (the finite element method).

The three steps of this explicit scheme enable us to get an error as small as $o({\Delta t}^3)$, the different steps are:

If $u_{n}=u(t)$ and $u_{n+1}=u(t+\Delta t)$, it leads to:

$$
u^\prime=u_{n}+\Delta t \gamma_{1}{\cal N}(u_{n})
$$
$$
u^{\prime\prime}=u^\prime+{\Delta t}[\gamma_{2}{\cal N}(u^\prime)+\zeta_{1}{\cal N}(u_{n})]
$$
$$
u_{n+1}=u^{\prime\prime}+{\Delta t}[\gamma_{3}{\cal N}(u^{\prime\prime})+\zeta_{2}{\cal N}(u^\prime)]
$$

The coefficients $\gamma_{1}$, $\gamma_{2}$, $\gamma_{3}$, $\zeta_{1}$ et $\zeta_{2}$ are deduced from the Taylor expansion of $u(t+\Delta t)$ and thus leads to:
$\gamma_{1}=8/15$, $\gamma_{2}=5/12$, $\gamma_{3}=3/4$, $\zeta_{1}=-17/60$ and $\zeta_{2}=-5/12$.

\subsection{Code validation}

The STELEM code was validated thanks to an international dynamo benchmark in Jouve et al. \cite{Jouve}. 

All data and notes can be found at the address: http://www.nordita.dk/~brandenb/tmp/benchmark 

\end{appendix}

\end{document}